\documentclass[reprint,amsmath,amssymb,twocolumn,notitlepage]{revtex4-2}
\usepackage{color}
\usepackage{graphicx}
\usepackage{dcolumn}
\usepackage{bm}

\begin{document}

\title{Composite colloidal assembly by critical Casimir forces}

\author{T.E. Kodger$^{\ast}$}
\affiliation{Van der Waals - Zeeman Institute of Physics, University of Amsterdam, Science Park 904, Amsterdam, The Netherlands}
\affiliation{Physical Chemistry and Soft Matter, Wageningen University and Research, Stippeneng 4, Wageningen, The Netherlands.}
\author{N. Farahmand Bafi$^{\ast}$} 
\affiliation{Institute of Physical Chemistry, Polish Academy of Sciences, Kasprzaka 44/52, PL-01-224 Warsaw, Poland.}
\author{M. Labb{\'e}-Laurent}
\affiliation{Max-Planck-Institut f{\"u}r Intelligente Systeme, Heisenbergstr. 3, D-70569 Stuttgart, Germany.}
\author{E. Steijlen}
\affiliation{Van der Waals - Zeeman Institute of Physics, University of Amsterdam, Science Park 904, Amsterdam, The Netherlands}
\author{A. Macio$\l{}$ek$^{\ast}$}
\affiliation{Institute of Physical Chemistry, Polish Academy of Sciences, Kasprzaka 44/52, PL-01-224 Warsaw, Poland.}
\affiliation{Max-Planck-Institut f{\"u}r Intelligente Systeme, Heisenbergstr. 3, D-70569 Stuttgart, Germany.}
\author{P. Schall$^{\ast}$}
\affiliation{Van der Waals - Zeeman Institute of Physics, University of Amsterdam, Science Park 904, Amsterdam, The Netherlands.\\}
\affiliation{$^{*}$ Denotes equal contribution.}

\begin{abstract}
We investigate the phase behaviour of mixtures of two populations of colloidal particles dispersed in a binary solvent system near its critical composition. The surfaces of particles are chemically modified to elicit a specific solvent affinity for one of the solvents. In this way, fluid-mediated interactions, which involve the critical Casimir effect, become particle population specific. As a result, the colloidal mixture shows a complex crystallization behavior reminiscent of the crystallization of atomic alloys. We show that the exquisite temperature dependence and reversibility of the critical Casimir interaction allows sampling the entire phase diagram of the binary system, and can be even used to anneal the crystalline microstructure analogous to temperature cycling of atomic alloy phases.
\end{abstract}

\maketitle

\section{Introduction}

\begin{figure}[b!]
\centering
  \includegraphics[width=0.9\columnwidth]{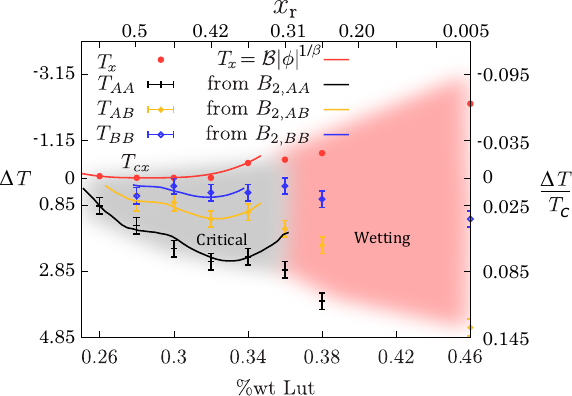}
  \caption{Temperature and solvent composition dependent aggregation diagram for a dilute composite system of colloidal particles A and B at equal particle number $N_\mathrm{A} = N_\mathrm{B}$.
The total volume fraction of colloids $\eta=\eta_A + \eta_B = 0.01$. Four aggregation boundaries are observed: $T_{AA}(c_L)$ marks the onset of  the self-aggregation for $P_\mathrm{A}$; $T_{AB}(c_L)$ the onset of  the hetero-aggregation for $P_\mathrm{A}$ and $P_\mathrm{B}$; $T_{BB}(c_L)$ the onset of the self-aggregation for $P_\mathrm{B}$. $T_{x}(c_L)$ is the solvent phase separation curve of the bulk solvent. The point $T_{cx}$ is the critical composition and temperature ($x_{r,c}$ = 0.5). Symbols correspond to the experimental data whereas solid lines are theoretical predictions (see SI for details of the theoretical model). From the fit of the solvent coexistence curve around the critical temperature of the solvent $T_\mathrm{c}$ to  $\phi=c_\mathrm{L}-c_\mathrm{L,c} = \mathcal{B}|t|^{\beta}$, where $t=\Delta T/T_\mathrm{c}=(T_\mathrm{c} - T)/T_\mathrm{c}$ and $\beta$ is the critical exponent, we have determined the non-universal amplitude $\mathcal{B}$.}
  \label{CC_phase_diagram}
\end{figure}

The formation of complex structures from micrometer- and nanometer-scale building blocks is currently of interest in bottom-up assembly. While the use of specific anisotropic shape can lead to potentially complex phases through volume exclusion~\cite{Glotzer2007} or directed interactions~\cite{Pine2012,Stuij2021a,Swinkels2021}, another route to realize complex structures, in particular, complex crystal structures, is through particle-specific pair interactions. In atomic systems, specific atomic pair interactions result from electronic shell structure that, together with quantum mechanical rules, ultimately determine the energy and coordination of the bonded state. In repulsive colloidal systems, binary crystal structures have been realized using particles of specific charge and size ratio, mimicking ionic crystals~\cite{Leunissen2005, Alsayed2005}. Recently, specific attractive interactions have been realized using so-called patchy particles~\cite{Feng2013}, together with DNA or critical Casimir-mediated interactions: in both cases, specific interactions of the patches with respect to the particle matrix have been obtained, which have enabled the assembly of simple molecule-like structures.

In particular, critical Casimir forces have been shown to provide reversible, temperature-dependent interactions that allow near-equilibrium assembly by taking advantage of the universal temperature dependence of the solvent correlation length setting the range and magnitude of the interaction~\cite{Maciolek2015}. Like the quantum-mechanical Casimir force, the critical Casimir force results from the spatial confinement of the fluctuating field, in this case the solvent composition between the particle surfaces close to the consolute point of a binary solvent.  In addition to the solvent correlation length, the interaction depends on the boundary conditions that define the confinement on the fluctuating solvent composition. Unlike the quantum-mechanical Casimir effect, in the critical Casimir analog, the boundary conditions can be easily tuned by controlling the wetting properties of the confining interfaces through surface treatment. Indeed, the recently realized patchy particle interactions take advantage of the surface-specific wetting properties to realize the interaction contrast between patch and matrix~\cite{Stuij2021a,Swinkels2021,Swinkels2024}. Doing this in a controlled way for mixtures of two population of colloidal particles, type A and B, would open the door to new binary particle structures mimicking complex phases of binary alloys in atomic systems. In those metallic alloy systems, annealing treatments are commonly applied to modify the crystalline microstructure through specific temperature cycling protocols, which can alter the grain size or defect concentration through thermal activation; a basic strategy in materials science. In colloidal structures, such annealing treatments have been poorly explored as typically the particle interactions are fixed, with temperature setting the thermal energy $k_\mathrm{B}T$, being a poor thermodynamic parameter. Yet, the use of reversible, temperature-dependent colloidal interactions would allow for heat treatments analogous to those of atomic systems, offering control over the crystallinity and microstructure that is central to many properties of crystalline materials.

In this paper, we demonstrate the formation of colloidal alloy phases due to particle-pair specific attractive interactions. The realised phases are reminiscent of atomic alloys, and result from critical Casimir interactions between two populations of surface-modified colloidal particles, A and B, that exhibit specific critical Casimir pair interactions. The specificity of  AA, AB, and BB  pair interactions is achieved by suitable \textit{surface} treatment of particles A and B such that they impose distinct boundary conditions on the concentration fluctuation of the solvent. We assemble colloidal binary liquid and crystal phases over a range of particle number ratios and attractions, exploring the full particle composition - temperature plane. To obtain insight into the phase behavior, we employ a mean-field model to calculate the full phase diagram for the binary colloidal mixture suspended in a  supercritical binary solvent. The model treats the solvent components explicitly, since for dense colloidal phases, the Casimir interactions are known to be highly non-additive~\cite{mattos2013many}.
We find that the topology of the phase diagram results from  competition of colloid-solvent and solvent-solvent couplings. Comparison of the predicted pair interactions and equilibrium phase diagram with the measurements shows that the observed colloidal phases are reminiscent of the crystallization of binary atomic alloys of particles with largely different condensation temperatures. Finally, we show that the reversibility of the interactions can be used to anneal the crystal microstructure using gentle temperature cycling. Our results of tunable, particle-pair specific colloidal interactions offer new opportunities for complex colloidal assembly, especially at the nanoscale, where large surface molecules, such as DNA, are prohibited.

\begin{figure*}[t!]
\centering
  \includegraphics[width=0.8\textwidth]{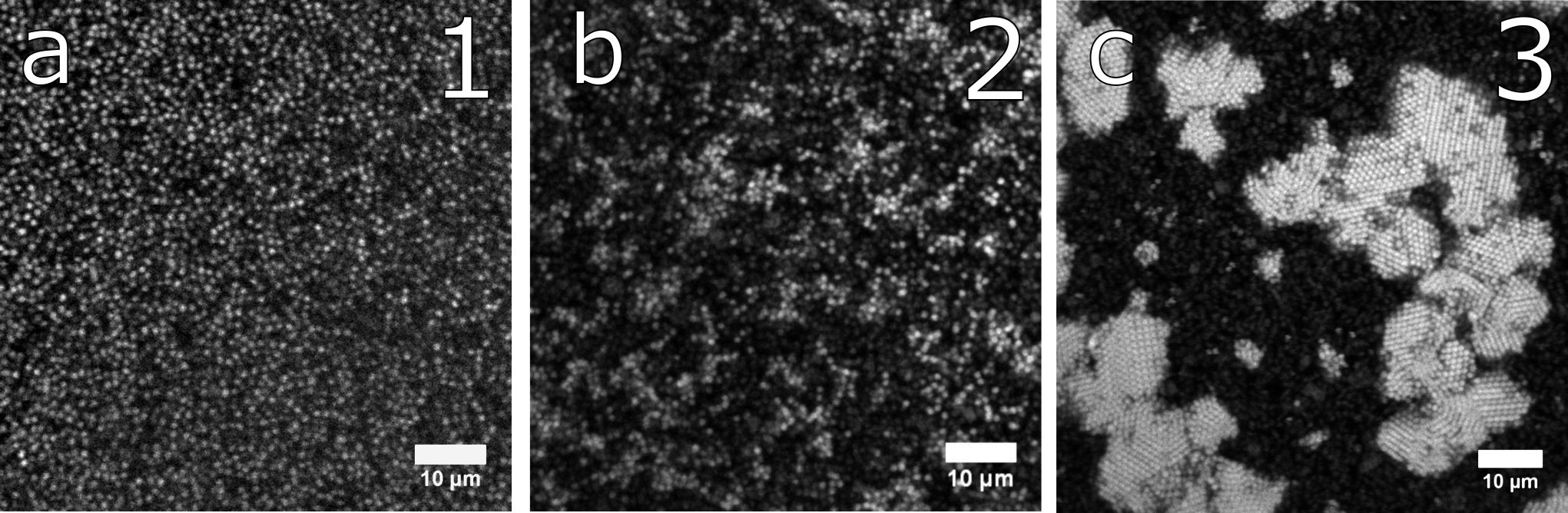}
  \caption{Temperature dependent $P_\mathrm{A}$ phase changes from colloidal gas, \#1, to colloidal liquid, \#2, to finally a colloidal crystal \#3 at $c_L = 32$wt\%. Note: upon $P_\mathrm{A}$ crystallite formation, sedimentation also occurs, resulting in an apparent compositional change.}
  \label{Temperature}
\end{figure*}

\begin{figure}[b!]
\centering
  \includegraphics[width=0.95\columnwidth]{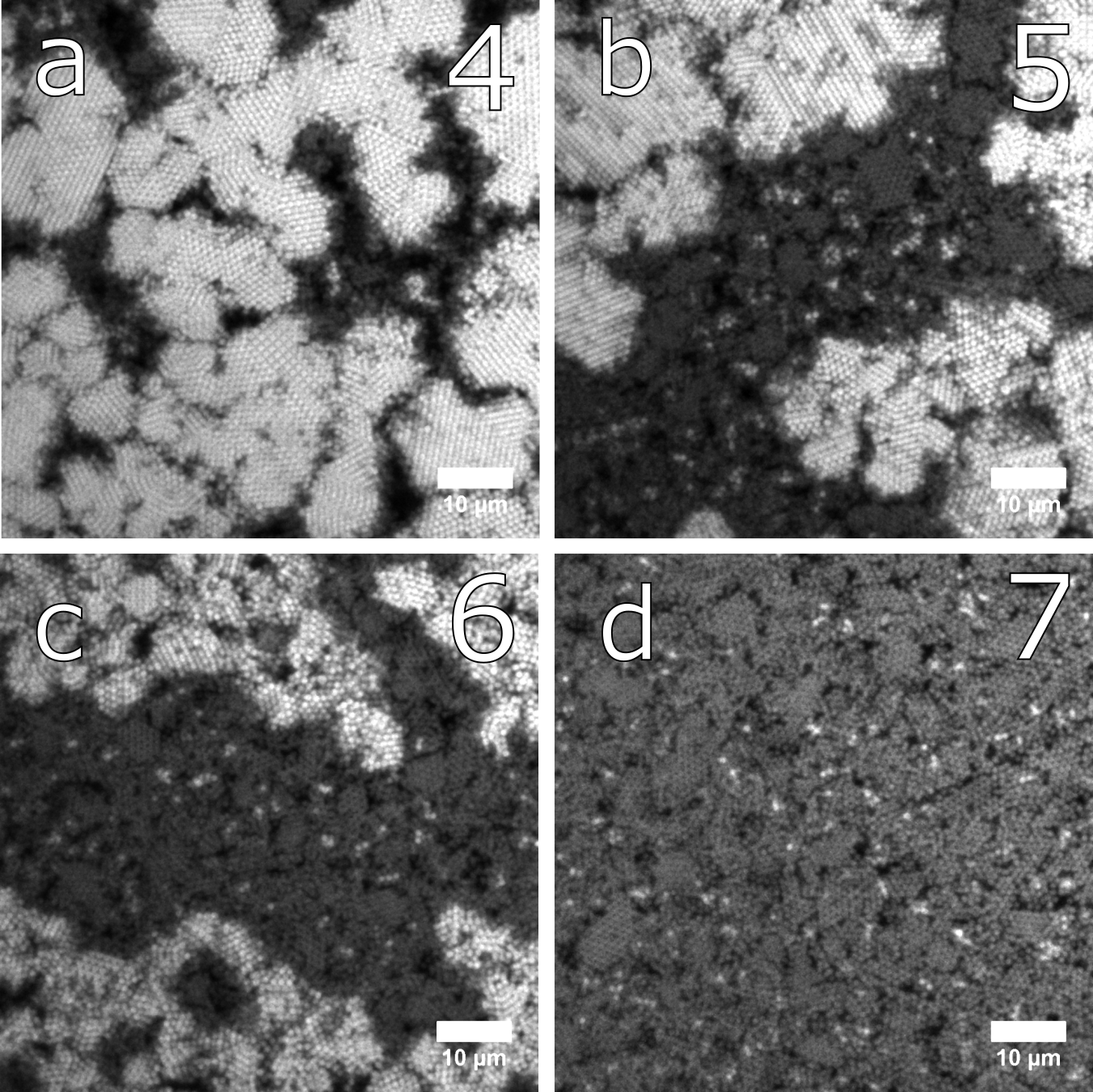}
  \caption{Compositionally dependent phases observed close to $T_\mathrm{c}$ at $\Delta T<T_\mathrm{BB}$ at  $c_\mathrm{L} = 32$wt\% corresponding to particle composition number indicated in the phase diagram shown in Fig.\ref{composition_phase_diagram}.}
  \label{Composition}
\end{figure}

\begin{figure}[b!]
\centering
  \includegraphics[width=0.9\columnwidth]{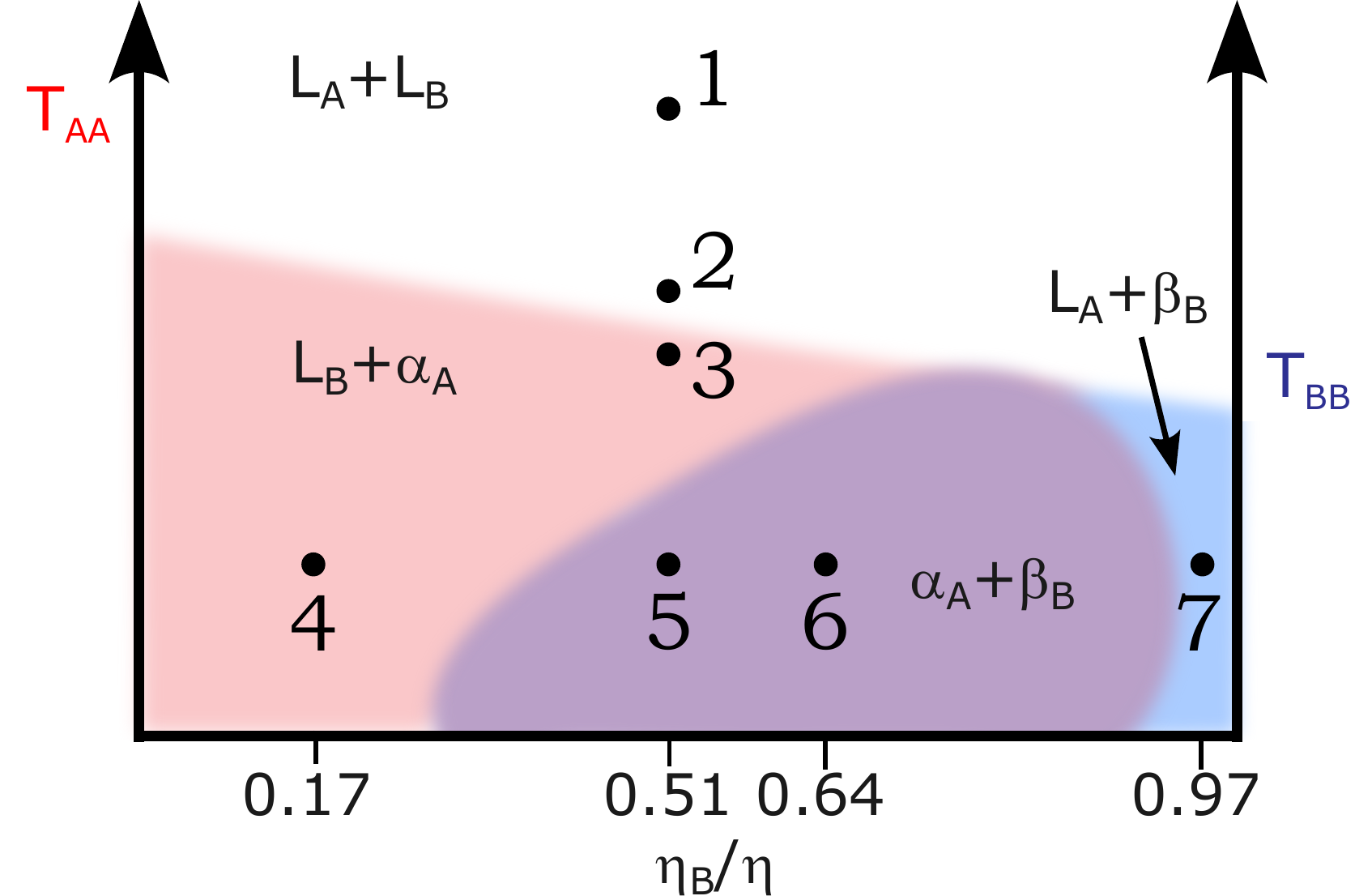}
  \caption{Temperature and particle composition ($\eta_B/\eta$) phase diagram with representative phase boundaries for all three composite phases: $L_{B}$+$\alpha_A$, $L_A$+$\beta_B$, and $\alpha_A$+$\beta_B$. Composition values are directly calculated by particle counting within images \#1-7 shown in Fig.~\ref{Temperature} and Fig.~\ref{Composition}.}
  \label{composition_phase_diagram}
\end{figure}

\section{Experimental Methods}
All chemicals were purchased from Sigma-Aldrich, and without further purification. We use poly(trifluoroethyl methacrylate) and poly(tert-butyl methacrylate) copolymer particles that can be easily surface modified due to an \textit{inimer} moiety on the surface~\cite{Kodger2015}. The particle batches A and B have nearly identical radii $R_\mathrm{A} = 510\,$nm and $R_\mathrm{B} = 475\,$nm, respectively, and identical low polydispersity, $\sigma/\langle r \rangle = 4 \%$, as determined by dynamic light scattering. We then alter the surface properties of particle A ($P_\mathrm{A}$) by surface-initiated atom transfer radical polymerization~\cite{Kodger2015}, rendering a surface charge density of $-0.09\,$e/nm$^{2}$ compared to unmodified particles B, $P_\mathrm{B}$, with $-0.15\,$e/nm$^{2}$. Note: Surface charge densities are calculated using $\zeta$-potential measurements linearly extrapolating to zero salt~\cite{Spruijt2011}, however, the presence of the extended surface-initiated brush prohibits accurate indication of surface charge densities. The magnitude of the surface treatment can be adjusted by altering the concentration of \textit{inimer} or the length of the grown polyelectrolyte from the surface~\cite{vanDoorn2017}; here, we use a short polyelectrolyte chain composed of a high stoichiometry of 2:1 charged monomer to neutral monomer (2-acrylamido-2-methyl-1-propanesulfonic acid sodium salt : N,N-dimethylacrylamide), to yield a strong contrast between modified particles A and unmodified B. Further details are provided in the Supplementary Information.

To visually distinguish $P_\mathrm{A}$ from $P_\mathrm{B}$, $P_\mathrm{A}$ are made with a 4X molar higher concentration of fluorophore, Cy3MM~\cite{Kodger2017}, thus under fluorescence imaging, they appear brighter and are distinguishable from $P_\mathrm{B}$.  Mixtures of particles A and B are then suspended in a binary solvent of 2,6-lutidine and deionized water with lutidine mass fraction $c_\mathrm{L}$ in the range of 26-46wt\%. The mixture of 2,6-lutidine and deionized water has a lower critical point at $T_\mathrm{cx}=32.85^{\circ}C$ and $c_\mathrm{L,c}\approx 0.286$.  A low amount of salt, 3mM NaCl, is added to fix the Debye screening length, $\kappa^{-1} \approx 3$nm; other sources of electrostatic impurities can therefore be neglected. To estimate the interaction contrast of the particles experimentally, we first determine the aggregation temperatures $T_{AA}, T_{AB}$ and $T_{BB}$  of particle pairs by direct observation of the onset of particle aggregation under the microscope for a dilute dispersion of particles with colloid volume fraction, $\eta=\eta_A + \eta_B = 0.01$. We then use the full temperature control of the particle pair interactions to explore the condensation and crystallization of particle mixtures across a wide range of particle number ratios, from A-rich to B-rich.

\section{Experimental Results}

Indeed, the surface-modified particles $P_\mathrm{A}$ aggregate at temperatures furthest away from $T_\mathrm{c}$, indicating that they exhibit the strongest critical Casimir attraction. The observed aggregation curves  $T_\mathrm{AA}$, $T_\mathrm{AB}$ and $T_\mathrm{BB}$ presented in Fig.~\ref{CC_phase_diagram} are clearly different for the three pairs, dropping from BB, which aggregate closest to $T_\mathrm{c}$ via AB to AA, which aggregates furthest away from $T_\mathrm{c}$. The critical Casimir interaction regime around $c_\mathrm{L} \sim 0.29 wt\%$ is clearly distinct from the wetting regime, characterized by much larger particle interactions, as confirmed by explicit modeling of the Casimir pair interactions. To compute the pair interactions in this dilute regime, we use a minimal model based on the sum of a screened electrostatic repulsion and a critical Casimir attraction. For the repulsive part of the potential, we employ the Yukawa potential~\cite{russel1991colloidal,hansen2000effective,barrat2003basic}, whereas the critical Casimir potential (CCP) is evaluated within the Derjaguin approximation from results for the slab geometry obtained using the local functional approach~\cite{borjan2008off,mohry2014critical}, see Supplemental Materials. We assume the onset of  aggregation to occur when the second virial coefficient $B_2^{ij}$ of particle pairs $ij$, reduced by that of a \textit{h}ard-\textit{s}phere reference system, $B_2^{ij}/B_2^{(hs)}$, is equal to the critical value $B_2^*=-1.2$ of the sticky sphere model, which described previous critical Casimir-induced aggregation very well~\cite{Stuij-et:2017}.

Using this condition, we find that the surface-modified particles $P_\mathrm{A}$ are strongly hydrophilic, and the unmodified particles $P_\mathrm{B}$ are less hydrophilic than A particles as seen in Fig. \ref{CC_phase_diagram} by the larger $\Delta T$ for $P_\mathrm{A}$ compared to $P_\mathrm{B}$ at $c_\mathrm{L} = 0.34$wt\%. The resulting fits of the aggregation boundaries agree well with the measured aggregation data for $T_\mathrm{AA}$, $T_\mathrm{AB}$, and $T_\mathrm{BB}$ up to the concentration of lutidine $c_\mathrm{L} = 0.34$wt\%, as shown in Fig. \ref{CC_phase_diagram}.
Due to the stronger preferential adsorption for water, the critical Casimir attraction for the AA pairs is stronger than for the BB pairs; at the same time the strength of their repulsion, which is proportional to the surface charge, is smaller, while its range, the Debye screening length $\kappa^{-1}$, is the same for all pairs. Both stronger attractive and weaker repulsive components increase the effective attraction, making the A particles aggregate further away from the critical temperature.
The fits are valid until lutidine concentrations $c_\mathrm{L} \sim 0.36$, delineating the critical Casimir regime clearly from the wetting-dominated regime at higher lutidine concentrations, where wetting-induced capillary interactions lead to much stronger attractive forces~\cite{Hertlein2008}, which we found to be not suitable for equilibrium assembly.

Using the particle-pair specific critical Casimir interactions, we explore the assembly of complex binary colloidal phases. We prepare particle mixtures first at equal concentrations of $P_\mathrm{A}$ and $P_\mathrm{B}$ with total volume fraction  $\eta= 0.4 \pm 0.04$ in solvents with  $c_\mathrm{L} = 32$wt\%, corresponding to the minimum of the aggregation temperatures in Fig. \ref{CC_phase_diagram}, and use temperature to control the strength and range of the critical Casimir pair interactions. Increasing the sample temperature towards $T_\mathrm{AA}$, we observe that the particles condense from a colloidal gas (Fig.~\ref{Temperature}a) to a colloidal liquid (Fig.~\ref{Temperature}b) and subsequently from colloidal liquid to crystal (Fig.~\ref{Temperature}c). These condensed phases consist mostly of A particles with only a small amount of particles B. Interestingly, the nucleating colloidal liquid phase appears rather ramified, suggesting that the liquid phase may be entered close to the gas-liquid critical point; see Supplemental Materials for a video of this condensed phase.
Upon further reducing $\Delta T$, particles of type B first assemble onto the already grown crystals of particles A at $T_\mathrm{AB}$ and then, upon reaching $T \sim T_\mathrm{BB}$, liquefy and crystallize between the A-rich crystals, as shown in Fig.~\ref{Composition}b. These observations are consistent with the particle pair interactions increasing from particle pairs BB to AB, and to AA. We thus obtain colloid A-rich crystals ($\alpha_\mathrm{A}$) coexisting with colloid B-rich crystals ($\beta_\mathrm{B}$) in a complex microstructure. Each crystal phase contains only a small fraction of the other component; the A-rich and B-rich crystals crystallize independent of each other due to the large temperature gap between $T_\mathrm{BB}$ and $T_\mathrm{AA}$ at $c_L = 32$wt\%.
To explore the binary colloidal phase diagram in more detail, we investigate particle mixtures from A-rich to strongly B-rich, as shown in Fig.~\ref{Composition} and the schematic diagram in Fig.~\ref{composition_phase_diagram}. We observe that  colloidal mixtures crystallize in a microstructure that consists of predominantly A-rich crystals for the A-rich mixture, and B-rich crystals for the  B-rich mixture, as shown in Fig.~\ref{Composition}. The separate $\alpha_\mathrm{A}$ and $\beta_\mathrm{B}$ crystals have only a small amount of particles of opposite type incorporated. This small degree of mixing becomes particularly apparent at equal or similar particle composition, i.e. ($\eta_\mathrm{B}/\eta$) $\approx$ 0.5. This result shows the strong energetic preference for forming AA particle pairs, dominating over the entropy of mixing free energy. Furthermore, the remaining size difference between particles A and B, with particles A having a $\sim 7\%$ larger diameter than particles B, causes significant strain and related strain energy when incorporated in the $\beta_\mathrm{B}$ lattice.
Nevertheless, at the B-richest composition, $\eta_\mathrm{B}/\eta$ = 0.97, a solid solution is formed, with A particles uniformly incorporated in the $\beta_\mathrm{B}$ crystals as seen in Fig.~\ref{Composition}d.

\section{Calculations of phase diagram}

To gain insight into the complex phase behavior of the binary colloidal system, we calculate the phase diagram of the colloidal suspension in the near-critical solvent mixture. Because in the dense suspension, many-body solvent mediated interactions between colloids might come to the fore, we consider an explicit four-component colloid-colloid-solvent-solvent (ABCD) mixture using an approach that does not rely on the assumption of effective potential pairs used in earlier  studies~\cite{mohry2012phase,dang2013,mohry2014critical}. Our theory is the simplest mean-field treatment of a four-component mixture where two  of the species (colloids A and B) are much larger in size than the other two components (solvent molecules, referred to as C and D), and where the large species has an inherent preference to adsorb one of the two smaller species (species D).  Using free-volume arguments, we write down a mean-field approximation for the Helmholtz free energy, which can be decomposed as $F_\mathrm{MF} = F_\mathrm{AB} + F_\mathrm{CD} + U_\mathrm{ABD}$. The colloid contribution $F_{AB}(\eta_\mathrm{A},\eta_\mathrm{B},T)$ depends on the colloidal volume fractions $\eta_\mathrm{A}$ and $\eta_\mathrm{B}$.  $F_\mathrm{CD}(x,\eta,T)$ is the mean-field free energy of the binary CD solvent mixture in the free space in between the colloids  with fraction $x/(1-\eta)$ for species $D$ (water) and fraction $1-x/(1-\eta)$) for species C (lutidine). Finally, $U_\mathrm{ABD}(x,\eta_\mathrm{A},\eta_\mathrm{B},T;s_\mathrm{A},s_\mathrm{B})$ is the average adsorption energy of solvent D on the surfaces of colloids A and B, where the parameters $s_i>0, i=$A,B  control the energy gain due to adsorption of D molecules on the colloid A and B. Mimicking the experimental situation, we assume  $s_\mathrm{A} > s_\mathrm{B}$. The expressions of all contributions as well as the chosen values of parameters are given in the Supplemental Materials. This model, although simple, contains all the necessary physics to describe the near-critical  solvent-mediated  interactions  between colloids. Since an explicit solvent is used, fractionation of the system and many-body effects arise naturally. We study this model in 3D. In the limit $\eta\to 0$, the ABCD model reduces to a binary (CD) solvent mixture, which exhibits phase separation into C-rich and D-rich phases ending at the critical temperature $T_c$. The thermodynamic state of the CD mixture is characterized by $\Delta T/T_\mathrm{c} =(T-T_\mathrm{c})/T_\mathrm{c}$  together with the (reduced) chemical potential difference $\Delta\mu_\mathrm{s}=(\mu_\mathrm{D}-\mu_\mathrm{C})$ between species D and C. For $\Delta\mu_\mathrm{s}<0$, the system is C-rich for all temperatures. We follow the experimental situation and study colloids immersed in a mixed (one-phase) CD mixture, poor in the colloid-preferred species D relative to the critical composition ($\Delta T >0$ and $\Delta\mu_\mathrm{s}\le 0$ ). In the limit  $\eta_\mathrm{A}\to 0$ (or $\eta_\mathrm{B}\to 0$) the ABCD mixture reduces to the ternary colloid-solvent-solvent ACD (or BCD) mixture, which was studied in two dimensions in Refs.\cite{edison2015critical,10.1063/1.4961437,edison2015phase} and in three dimensions in Ref.~\cite{10.1063/1.4979518}. Here, we extend the system to two types of colloids to model particle A-B mixtures:  Assuming that they have the same size, but their interaction with the solvent D is different, we find stable colloidal gas, liquid, and crystal phases, as well as broad coexistence of gas-liquid and gas-crystal phases and clear fractionation of the solvent in the coexisting colloidal phases.

\begin{figure}[t!]
\centering
      \includegraphics[width=0.8\columnwidth]{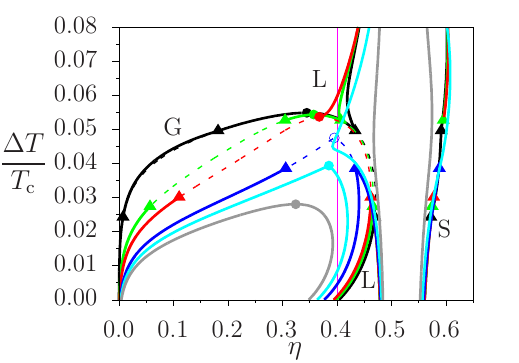}
      \includegraphics[width=0.8\columnwidth]{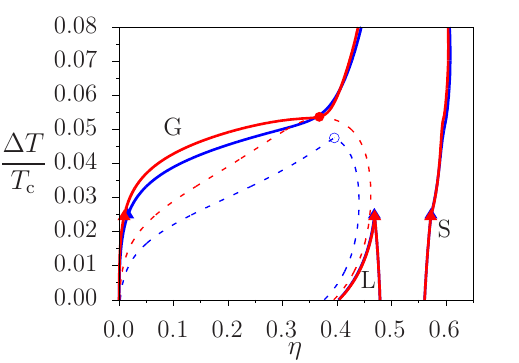}
\caption{
Theoretical phase diagram of an ABCD mixture in the plane $(\Delta T/T_c, \eta)$ for a pure solvent composition of $x_\mathrm{r}=0.4$, representing the gas (G), liquid (L), and solid (S) phases. (a) Color coding indicates different fixed fractions of $\eta_\mathrm{B}/\eta$ in the \textit{liquid} phase: black, green, red, blue, cyan, and gray for $\eta_B/\eta=$0, 0.1, 0.187, 0.5, 0.73, and 1, respectively. Full circles indicate stable upper critical points, while full triangles indicate triple points. Metastable G-L phase boundaries are indicated by dashed lines. For equal fractions of both colloid types (blue line), the upper critical point is metastable (open blue circle). The vertical magenta line shows the value of the experimental significant total packing fraction $\eta=0.4$ for which we present the phase behavior in Fig.~\ref{fig:mf_2}. (b) The phase diagram for the fraction $\eta_\mathrm{B}/\eta$ fixed in the \textit{gas} phase  at the value of 0.187 (red line) and 0.5 (blue line) showing broad coexistence of G-S phases. Metastable G-L phase boundaries with the same value of $\eta_\mathrm{B}/\eta$ are indicated by dashed lines. Note that the triple points differ from those in (a), where the $\eta_\mathrm{B}/\eta$ ratio is fixed in the liquid phase.
}
  \label{fig:mf_1}
\end{figure}

The resulting phase diagram in the $(\Delta T/T_\mathrm{c},\eta)$ plane is shown in Fig.~\ref{fig:mf_1}(a); it is computed  for a fixed value of the pure solvent concentration $x_\mathrm{r}=0.4$, slightly deviating from its critical value $x_{\mathrm{r,c}}=0.5$, similar to the experiment. (Note that the composition x of the solvent in the coexisting phases is generally very different from the composition $x_\mathrm{r}$ of the solvent reservoir.)  The color corresponds to various values of colloid B fraction $\eta_\mathrm{B}/\eta$, which is fixed  in the liquid phase, as discussed in Fig.~\ref{fig:mf_1} caption.  The topology of the phase diagram is very sensitive to the value of $\eta_\mathrm{B}/\eta$.  For small fractions of the less adsorbing colloid type, we observe two regions of stable gas-liquid (G-L) coexistence: (1) a tiny upper G-L coexistence at higher temperatures  terminating at the top with the upper critical point and at the bottom with the upper triple point, in which the gas, liquid and solid phases coexist; (2) a wide lower G-L coexistence closer to $\Delta T/T_\mathrm{c}=0$ terminating at the top with the lower triple point. Note: The lower critical point lies below the critical temperature of the solvent and is therefore not shown here. The upper G-L coexistence narrows with increasing $\eta_\mathrm{B}$ fraction and disappears altogether at $\eta_\mathrm{B}/\eta\simeq 0.187$ as shown as the red line in Fig.~\ref{fig:mf_1}. There remains a wide lower G-L coexistence ending at the triple point, above which the G-L coexistence becomes metastable with respect to the gas-solid (G-S)  coexistence (see Fig.~\ref{fig:mf_1}(b))\footnote{The reason the G-S coexistence is not depicted in Fig.~\ref{fig:mf_1}(a) is that there we set $\eta_\mathrm{B}/\eta$ in the liquid phase, which disappears once $\eta_\mathrm{B}/\eta\simeq 0.187$ is reached. As a result, only G-S coexistence occurs  at high temperatures. Fixing $\eta_\mathrm{B}/\eta$ in the gas phase above $\eta_\mathrm{B}/\eta\simeq 0.187$ causes G-S to terminate in a new set of triple points, different from those in  Fig.~\ref{fig:mf_1}(a). This is shown in Fig.~\ref{fig:mf_1}(b) where we fix the value $\eta_\mathrm{B}/\eta$ in the gas phase with the same color code as in the  panel (a). As can be seen, the L-G transition with the same value of $\eta_\mathrm{B}/\eta$ in the liquid phase (which is the same as in the panel (a)) is metastable with respect to the G-S transition. The same happens for the G-S transition between the triple points in the green curve (not shown in the panel (b)).}.
Upon adding  even more B-type colloids into the dispersion, the G-S coexistence shrinks and moves towards higher temperatures with lower triple points moving upwards, as seen in Fig.~\ref{fig:mf_1}(a). Finally, for $\eta_\mathrm{B}/\eta \simeq 0.689$, the topology of the phase diagram changes: the triple points converge to a single point, and a stable upper critical point reappears. This topology remains until the mixture contains only B-type colloids (cyan and grey lines in Fig.~\ref{fig:mf_1}(a)). We observe that the liquid (L) - solid (S) transition curves remain fairly unchanged with changing $\eta_\mathrm{B}/\eta$, as they are mainly dictated by packing considerations, closely resembling those of a monodisperse hard-sphere system.

\begin{figure}[t]
\centering
  \includegraphics[width=0.95\columnwidth,trim={0 0 0 5cm},clip]{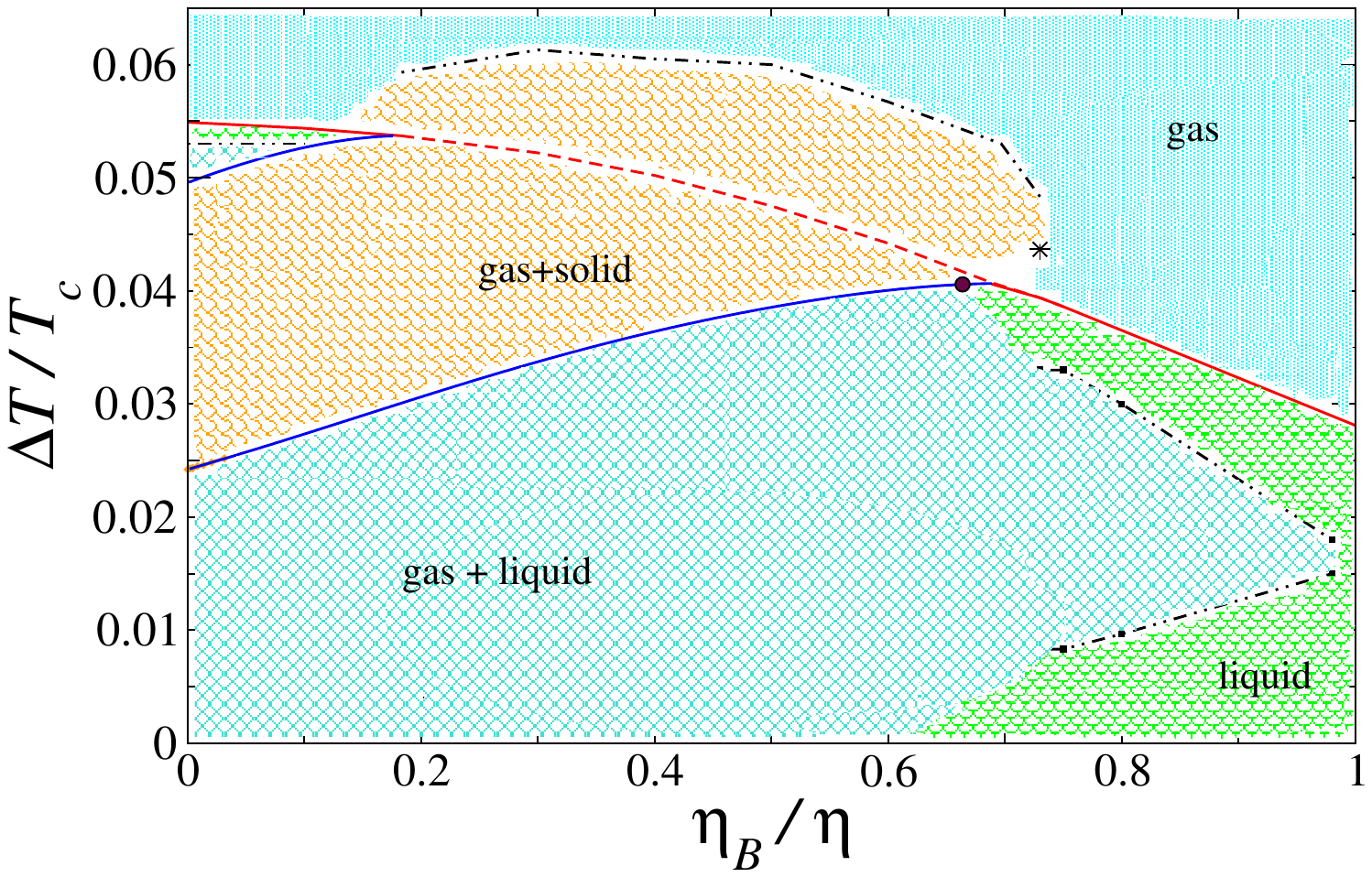}
  \caption{Theoretical phase behavior in the temperature-colloidal composition plane $\eta_\mathrm{B}/\eta$ for a pure solvent composition of $x_\mathrm{r}=0.4$ and a fixed value of the total colloid packing fraction $\eta=0.4$. The upper critical point line is marked in red -- its dashed portion corresponds to metastable critical points. The liquid branches of the upper and lower triple points are shown in blue. The triple point lines terminate at the critical line. The dashed lines indicate the gas-solid and liquid-gas phase boundaries. They correspond to the intersections of the vertical magenta line with the various phase boundaries in Fig. ~\ref{fig:mf_1}.}
  \label{fig:mf_2}
\end{figure}

We focus on the total colloid fraction $\eta = 0.4$, magenta vertical line in Fig.~\ref{fig:mf_1}, which is close to the experimental value, and show the phase behavior as a function of colloid B fraction in Fig.~\ref{fig:mf_2}. For this value of $\eta$, gas-solid coexistence is observed already for large $\Delta T$, as the gas-solid lines in Fig.~\ref{fig:mf_1}(a) are bent towards low $\eta$, shifting them into the $\eta = 0.4$ plane. 
The metastable gas-liquid coexistence is observed at slightly lower $\Delta T$, demarcated by the dashed red line, with stable gas-liquid coexistence at low and high $\eta_\mathrm{B}$ (solid red line).

Close to $T_\mathrm{c}$, we observe only a liquid phase or liquid-gas coexistence, while liquid-solid coexistence would be observed at higher $\eta$, outside the current $\eta = 0.4$ plane. In a typical route upon cooling, one would first encounter gas-solid or gas-liquid-solid transitions, separating the system into low-$\eta$ gas and high-$\eta$ solid regions, which subsequently would remain stable towards lower $\Delta T$. Our experimental observations above are consistent with these predictions: The predicted metastable gas-liquid transition and critical point are consistent with the observed gas-liquid density fluctuations in Fig.~\ref{Temperature}b, and the predicted underlying gas-solid coexistence is in agreement with the stable gas-solid coexistence observed some time later, at slightly lower $\Delta T$, see Fig. ~\ref{Temperature}c.
Eventually, the remaining B-rich gas phase crystallizes to form the B-rich crystallites visible in Fig.~\ref{Composition}b. It should be emphasized that although our model accounts for the colloid adsorption contrast and their influence on the phase behavior, it cannot distinguish the structure of the solid phase, e.g. which type of colloidal particles constitute the crystal at a given temperature. To predict them, other approaches will be necessary, e.g., Monte Carlo simulations analogous to those in Refs~\cite{edison2015critical,10.1063/1.4961437,10.1063/1.4979518}
but extended to four-component mixtures, or molecular dynamics simulations~\cite{dang2013,PhysRevE.59.5744}. However, the latter are only feasible for colloidal mixtures in an \textit{implicit} solvent and are based on the effective potential of a single colloid pair, thus neglecting many-body interactions, which become increasingly important in the case of dense suspensions near the solvent critical point.

\section{Annealing of crystal phases}

\begin{figure}[t]
\centering
  \includegraphics[width=0.9\columnwidth]{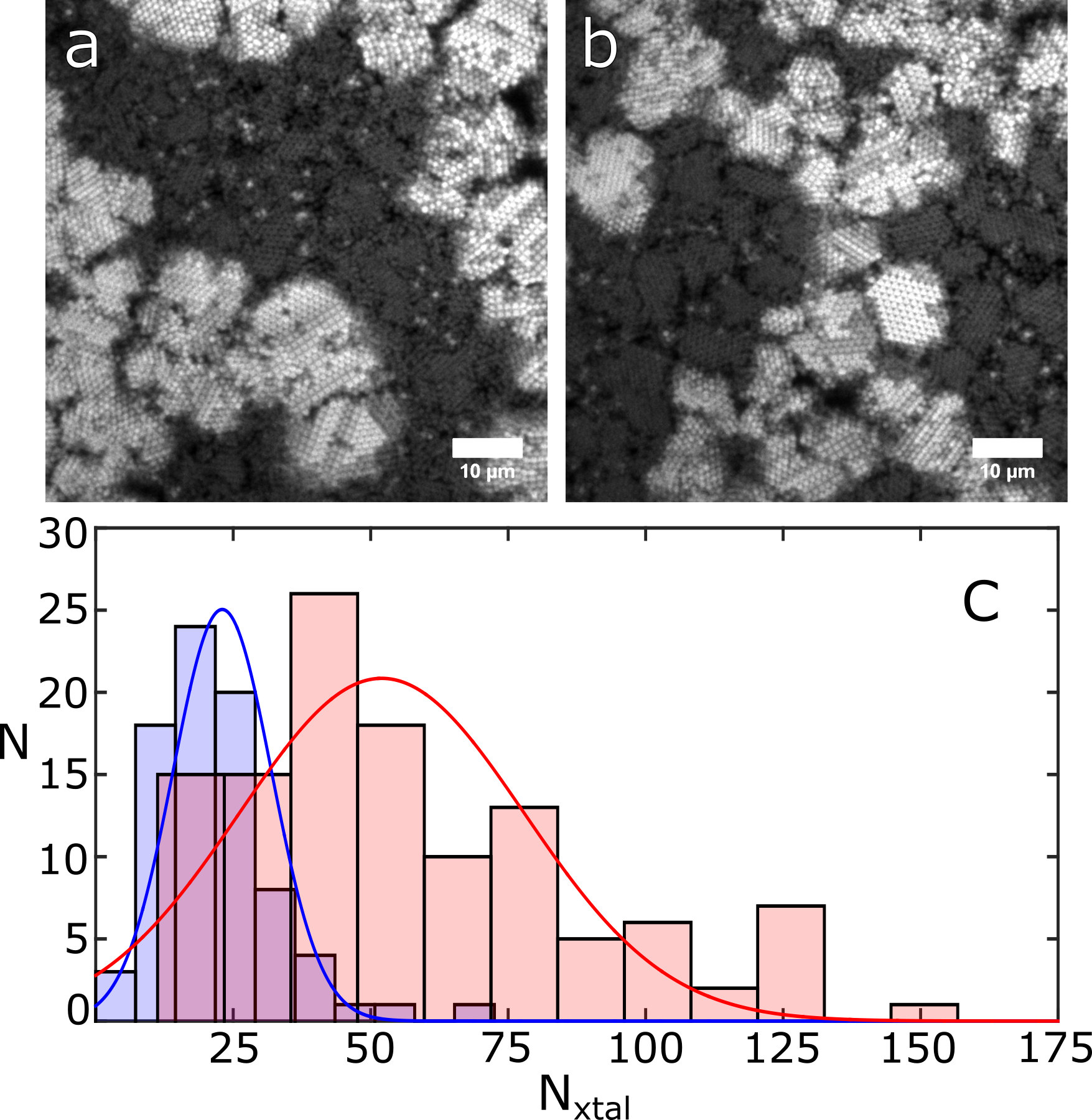}
  \caption{Sample with composition $\eta_\mathrm{B}/\eta=0.51$ (sample 5, Fig.~\ref{composition_phase_diagram}) (a) before and (b) after thermal annealing by reducing $T= 32.41^{\circ}$C $= T_\mathrm{BB}-0.01^{\circ}C$ for 4hrs, then increasing $T=T_\mathrm{BB}=32.41^{\circ}$C. (c) Distribution of the frequency ($N$) of a given $P_\mathrm{B}$ crystallite size ($N_{\chi}(P_\mathrm{B})$) for the sample before (blue) and after annealing (red); solid lines are Gaussian fits to distribution.}
  \label{annealing}
\end{figure}

A key advantage of the critical Casimir interaction is the use of temperature to dynamically control the interparticle potential. This control may permit the annealing of the crystalline microstructure by cycling the particle interactions \textit{in situ}. Indeed, we observe that particle condensation and crystallization are fully reversible: The assembled crystals melt and dissociate when we lower the temperature below the respective thresholds $T_\mathrm{BB}$ and $T_\mathrm{AA}$, see Supplemental Materials for video. This reversibility opens the opportunity to use temperature cycling to ripen the crystalline microstructure. To explore the effect of such annealing, starting near $\Delta T = 0.43^{\circ}$C, which is within $0.02^{\circ}$C but above $T_\mathrm{BB}$, we reduce the temperature by $0.01^{\circ}$C for 4 hours, and subsequently increase it back to  $\Delta T = 0.43^{\circ}$C. Indeed, we observe that the resulting crystallinity of the B-rich domains increases as shown by the microscope images in Fig.~\ref{annealing}a and Fig.~\ref{annealing}b taken before and after annealing, respectively. The sizes of the B-rich crystals grow, as demonstrated by the crystal size distribution in Fig.~\ref{annealing}c that shifts significantly to larger crystal sizes upon annealing. We can thus alter the microstructure of the binary colloidal crystals by annealing, in analogy to the annealing treatments of metallic alloys; the latter is essential for adjusting the microstructure to the most favorable structure for applications.  Similarly, in the binary colloidal crystals, the reversible critical Casimir interaction allows tuning the colloidal microstructure; repeating this annealing step with optimized temperature step should allow for further ripening the B-rich crystals to large crystal size.
Finally, we note that the absorption contrast between colloids type A and B can be tuned through surface charge. This allows adjusting the binary colloid phase diagram to explore more binary alloys properties with this colloidal model system.

\section{Conclusions}
Surface-modified colloidal particles giving rise to specific critical Casimir pair interactions in near-critical solvents offer new opportunities for the assembly of complex crystalline phases. The surface modification defines the strength of the attractive critical Casimir interaction; using particles of types A and B with different hydrophilic character, we obtain attractive interactions of strength $u_\mathrm{AA}$, $u_\mathrm{AB}$, and $u_\mathrm{BB}$ that give rise to crystalline phases reminiscent of binary alloy phases of atoms with different condensation temperatures. We have shown that a minimal mean-field explicit four-component model captures the underlying phase behavior of the experimental system, while also predicting additional critical and triple points that have yet to be realised. Future systems with more similar surface affinities and thus closer interaction strengths will allow the  assembly of solid solutions of A and B, as analogs of eutectic metallic alloy phases. Additionally, exploring other $c_\mathrm{L}$ with smaller temperature gaps between $T_\mathrm{AA}$ and $T_\mathrm{BB}$ may offer a unique colloidal methodology to tune the condensation temperature difference between the two particles. The unique temperature control of the pair interactions allows tuning the crystalline microstructure as we have shown by temperature cycling to enhance the grain size, in analogy to annealing treatments of metallic alloys.

\section{Acknowledgments}
T.E.K. and P.S. acknowledge funding by a VICI grant (grant nr. 680.47.615) of the Netherlands Organization for Scientific Research (NWO). The research of N.F.B and A.M. was funded by  the  National Science Center, Poland (Opus Grant  No.~2022/45/B/ST3/00936). N.F.B. and A.M. thank R. Evans for stimulating discussions regarding the formulation of the four-component mixture model.

\bibliography{references}

\end{document}